\documentclass[runningheads]{svmult}

\usepackage{graphicx}  % standard LaTeX graphics tool
\newcommand{\eq}[1]{(\ref{#1})}

\title*{Lamb Shift in Light Hydrogen-Like Atoms}

\author{
Vladimir G. Ivanov\inst{1}\thanks{E-mail: ivanovv@gao.spb.ru}
\and Savely G. Karshenboim\inst{2,3}\thanks{E-mail: sek@mpq.mpg.de}
}

\institute{
Pulkovo Observatory, St. Petersburg, Russia
\and D. I. Mendeleev Institute for Metrology, St. Petersburg, Russia
\and Max-Planck-Institut f\"ur Quantenoptik, Garching, Germany\thanks{The summer address}
}

\begin{document}

\maketitle

\begin{abstract}
Calculation of higher-order two-loop corrections is now a limiting factor
in development of the bound state QED theory of the Lamb shift in the hydrogen atom
and in precision determination of the Rydberg constant. Progress in the study of light
hydrogen-like ions of helium and nitrogen can be helpful to investigate these uncalculated
terms experimentally. To do that it is necessary to develop a theory of such ions. We
present here a theoretical calculation for low energy levels of helium and nitrogen ions.
\end{abstract}

\section{Introduction}
The Quantum Electrodynamics (QED) theory of simple atoms like hydrogen or
hydrogen-like ions provides precise
predictions for different energy levels \cite{icap,mplp}. Particularly,
some accurate results were obtained
for the Lamb shift in the ground state of the hydrogen atom. The accuracy of the QED
calculations of the Lamb shift has been limited by unknown higher-order two-loop corrections
and inaccuracy of determination of the proton charge radius \cite{karshenboim99}.
As far as the proton size
is going to be determined very precisely from a new experiment \cite{pohl}
on the Lamb shift in muonic
hydrogen, the only theoretical uncertainty is now due to the two-loop contribution.
Improvement of the theory is important to determine the Rydberg constant with high accuracy
\cite{twh,schwob} and to test the bound state QED precisely.

Since the theory seems not to be able to give now any results on higher-order two-loop
corrections ($\alpha^2(Z\alpha)^6m$ and higher) we have to look for another way to estimate
these terms and so the uncertainty of the hydrogen Lamb shift theory. An opportunity is to
study the problem experimentally, measuring the Lamb shift in different hydrogen-like ions
at not too high value of the nuclear charge $Z$. Only for two such ions the Lamb shift can
be available with a high accuracy from experiment at the present time or in the near future. Namely,
these are helium
\cite{drake,burrows} and nitrogen \cite{myers} ions.
The experimental estimation of higher-order two-loop terms is quite of interest also
because of recent speculation on a great higher-order term \cite{mallam} (see also
Refs. \cite{goidenko,yerokhin}).

The advantage of using $Z>1$ is determined by the scaling behaviour of different QED values:
\begin{itemize}
\item The scaling of the Lamb shift is $Z^4$;
\item The scaling of the radiative line width of excited states
(e. g. of the $2p$ and $3s$ states) is $Z^4$ as well;
\item The scaling of the unknown
higher-order two-loop corrections to the Lamb shift is $Z^6$.
\end{itemize}
Thus, relatively imprecise measurements with higher $Z$ can nevertheless give some
quite accurate data on some QED corrections.

Our target is to develop a theory for the Lamb shift and the fine structure in these two
atomic systems. Eventually we need to determine the $2s-2p_{1/2}$ splitting
in the helium ion (for comparison with
the experiment
\cite{drake}), difference of the Lamb shifts $E_L(2s)-E_L(3s)$ in $^4$He$^+$ (for the  project
\cite{burrows}) and the $2p_{3/2}-2s$ interval in hydrogen-like nitrogen. The difference
mentioned is necessary \cite{zp97,icap} if one needs to compare the results of the Lamb shift
($n=2$) measurement \cite{drake} and the $2s-3s$ experiment.

Since the uncertainty of the QED calculations is determined for these two
ions (He$^+$ and N$^{6+}$) by the higher-order two-loop terms, we are going to reduce
the other sources of
uncertainty. We present results appropriate to provide an interpretation of the experiments
mentioned as a direct study of the higher-order two loop corrections. The results
of the ions experiments should afterwards be useful for the hydrogen atom.

\section{Theoretical contributions}

\subsection{Definitions and notation}

The Lamb shift is defined throughout the paper
as a deviation from an unperturbed energy level\footnote{We use the
relativistic units in which $\hbar=c=1$.}
\begin{equation} \label{mMDir}
 E^{(0)}(nl_j) = m_R\Bigl[f(nj) - 1\Bigr] -\frac{m_R^2}{2(M+m)}\Bigl[f(nj) -1\Bigr]^2
 \,,
\end{equation}
where $M$ and $m$ are the mass of the nucleus and of the electron, and $m_R$ stands for the reduced
mass.  The dimensionless Dirac energy of
the electron in the external Coulomb field is of the form
\begin{equation}
 f(nj) = \left(1+\frac{(Z\alpha)^2}{\left(n-j-\frac{1}{2}+
 \sqrt{(j+\frac{1}{2})^2-(Z\alpha)^2}\;\right)^2}\right)^{-\frac{1}{2}}
 \,.
\end{equation}

The Lamb shift is mainly a QED effect, perturbed by the influence of
the nuclear structure
\begin{equation}
 \Delta E = \Delta E_{\rm QED} + \Delta E_{\rm Nucl}
 \,.
\end{equation}
The QED contribution
\begin{equation}
 \Delta E_{\rm QED} = \Delta E_\infty + \Delta E_M
\end{equation}
includes one-, two- and three-loop terms calculated
within the external-field approximation
\begin{equation}
 \Delta E_\infty =
 \frac{\alpha(Z\alpha)^4}{\pi} \left( \frac{m_R}{m} \right )^3
 \frac{m}{n^3}
 \left[
  F + \left(\frac{\alpha}{\pi}\right) H + \left(\frac{\alpha}{\pi}\right)^2 K
 \right]\,,
\end{equation}
and a recoil corrections
$\Delta E_M$, which is a sum of pure recoil and radiative recoil
contributions depending on the nuclear mass $M$
\begin{equation}
 \Delta E_{M} = \Delta E_{\rm Rec} + \Delta E_{\rm RRC}
 \,.
\end{equation}

\subsection{One-loop contributions: self energy of the electron}

Let us start with the one-loop contribution.
The terms of the external-field contribution are usually written in the
form of an expansion
\begin{equation} \label{exp}
 (Z\alpha)^4 F(Z)
 =
 \sum_{i,j}  A_{ij}
 (Z\alpha)^i \ln^j \frac{1}{(Z\alpha)^2}
 \,.
\end{equation}
The dominant contribution comes from the one-loop
self energy of the electron. The known results for some low levels are summarized
below\footnote{It is useful to keep somewhere the reduced mass $m_R$.}:
\begin{eqnarray}
  F^{SE}_{ns}(Z) = &~&
  \frac{4}{3} \ln \frac{m}{(Z\alpha)^2 \, m_R}
  - \frac{4}{3} \ln\big(k_0(ns)\big) + \frac{10}{9}
  \nonumber\\
  &+& (Z\alpha)\, 4\pi \left( \frac{139}{128} - \frac{1}{2} \ln(2) \right)
  + (Z\alpha)^2 \, \left(
    - \ln^2 \frac{1}{(Z\alpha)^2 }
    \right.
  \nonumber\\
    &+& \left.
      A_{61}(ns) \, \ln \frac{1}{(Z\alpha)^2 }
    + G_{ns}(Z)
    \right)
  \,, \\
  F^{SE}_{2p_{1/2}}(Z) = &  - &
  \frac{4}{3} \ln\big(k_0(2p)\big) - \frac{1}{6} \frac{m}{m_R}
  + (Z\alpha)^2 \, \left(
     \frac{103}{180} \, \ln \frac{1}{(Z\alpha)^2 }
     \right.
  \nonumber\\
    &+&
    \left. \vphantom{\frac11}
    G_{2p_{1/2}}(Z)
  \right)
  \,, \\
  F^{SE}_{2p_{3/2}}(Z) =
  & - & \frac{4}{3} \ln\big(k_0(2p)\big) + \frac{1}{12} \frac{m}{m_R}
  + (Z\alpha)^2 \, \left(
     \frac{29}{90} \, \ln \frac{1}{(Z\alpha)^2 }
     \right.
  \nonumber\\
    &+&
    \left. \vphantom{\frac11}
    G_{2p_{3/2}}(Z)
  \right)
  \,,
\end{eqnarray}
where the state-dependent logarithmic coefficient $A_{61}(ns)$ is known
\begin{eqnarray}
  A_{61}(1s) &=& \frac{28}{3}\ln(2) - \frac{21}{20} \,, \nonumber\\
  A_{61}(2s) &=& \frac{16}{3}\ln(2) + \frac{67}{30} \,, \nonumber\\
  A_{61}(3s) &=&  -4\ln(3) + \frac{28}{3} \ln(2) + \frac{6163}{1620} \nonumber\,,
\end{eqnarray}
the Bethe logarithm is \cite{bethe}
\begin{eqnarray}
 \ln\big(k_0(1s)\big) &=& 2.984\,128\,56\dots \,,\nonumber \\
 \ln\big(k_0(2s)\big) &=& 2.811\,769\,89\dots \,, \nonumber\\
  \ln\big(k_0(3s)\big) &=& 2.767\,663\,61\dots \,, \nonumber\\
  \ln\big(k_0(2p)\big) &=& -0.030\,016\,71\dots\nonumber
\end{eqnarray}
and higher-order self-energy terms $G_{nl}(Z)$ are numerically found in Sect. 2.5.

\subsection{One-loop contributions: polarization of vacuum}

The coefficient of the expansion \eq{exp} for free vacuum polarization can be calculated
in any order of $Z\alpha$ in a closed analytic form \cite{kars981}. In particular,
the result was found in Ref. \cite{kars981} for the circular states
($l=n-1$) with $j=l+1/2$.
For $n=1,2$ one can expand at low $Z\alpha$ and find
\begin{eqnarray}
\label{relvp1}
  F^{VP}_{1s}(Z) =&  - & \frac{4}{15}
  + \frac{5\pi}{48}(Z\alpha) \nonumber \\
  &+& (Z\alpha)^2 \, \left(
      - \frac{2}{15} \ln \frac{1}{(Z\alpha)^2 }
      + \frac{4}{15} \log(2)
      - \frac{1289}{1575}
    \right) \nonumber\\
  &+& (Z\alpha)^3 \left(
      \frac{5\pi}{96} \ln \frac{1}{(Z\alpha)^2 }
      + \frac{5\pi}{48} \ln(2) + \frac{23\pi}{288}
    \right) \,, \\
\label{relvp2}
  F^{VP}_{2p_{3/2}}(Z)=&-&
    \frac{1}{70} \, (Z\alpha)^2+ \frac{7\pi}{1024} \, (Z\alpha)^3 \,.
\end{eqnarray}

In the case of other states the result has been known only up to the order $(Z\alpha)^2$
\cite{mana,iknp,zp97}.
We present here new results for two other states at $n=2$ and for the $3s$ state:
\begin{eqnarray}
\label{relvp3}
  F^{VP}_{2s}(Z) =&-& \frac{4}{15}
  + \frac{5\pi}{48}(Z\alpha) \nonumber \\
  &+& (Z\alpha)^2 \, \left(
      - \frac{2}{15} \ln \frac{1}{(Z\alpha)^2 }
      - \frac{743}{900}
    \right) \nonumber\\
  &+& (Z\alpha)^3 \left(
      \frac{5\pi}{96} \ln \frac{1}{(Z\alpha)^2 }
      + \frac{5\pi}{24} \ln(2) + \frac{841\pi}{9216}
    \right) \,, \\
\label{relvp4}
  F^{VP}_{2p_{1/2}}(Z) =&-&
     \frac{9}{140} \, (Z\alpha)^2+\frac{41\pi}{3072} \, (Z\alpha)^3 \,, \\
\label{relvp5}
  F^{VP}_{3s}(Z) =&-&\frac{4}{15}
  + \frac{5\pi}{48}(Z\alpha) \nonumber \\
  &+& (Z\alpha)^2 \, \left(
      - \frac{2}{15} \ln \frac{1}{(Z\alpha)^2 }
      -\frac{4}{15} \ln \frac{3}{2} - \frac{1139}{1575}
    \right) \nonumber\\
  &+& (Z\alpha)^3 \left(
      \frac{5\pi}{96} \ln \frac{1}{(Z\alpha)^2 }
      + \frac{5\pi}{48} \ln(6) + \frac{137\pi}{2592}
    \right) \,.
\end{eqnarray}

\subsection{Wichmann-Kroll contributions}

It is not enough to consider the free vacuum polarization. The relativistic corrections
to the free vacuum polarization in Eqs. (\ref{relvp1}--\ref{relvp4}) are of the same order
as the so-called Wichmann-Kroll term due to Coulomb effects inside the electronic
vacuum-polarization loop.
To estimate this term we fitted
its numerical values from Ref. \cite{john}, which are more accurate for some higher $Z\sim30$,
by expression \cite{john,mana}
\begin{equation}
F_{WK}(Z) =  (Z\alpha)^2 \, \left( \frac{19}{45} - \frac{\pi^2}{27} \right)
  + (Z\alpha)^3 \, \left( A_{71} \ln\frac{1}{(Z\alpha)^2} + A_{70} \right)
  \,.
\end{equation}
We found that the contribution of $A_{70}$ and $A_{71}$ terms is small enough. The accuracy
of the calculation in Ref. \cite{john} is not high at $Z=7$ and we performed some fitting of higher $Z$
data. To make a conservative estimate we find two pairs of coefficients which reproduce the result
at $Z=30$. The results are:
\begin{eqnarray}
  A_{71}(ns) &=& -0.23(2), \quad  A_{70}(ns) = ~0 \quad {\rm and} \quad F_{WK}(7) = 0.000\;139(1)\;; \nonumber\\
  A_{71}(ns) &=& ~0,  \quad A_{70}(ns) = -0.07(1) \quad {\rm and} \quad F_{WK}(7) = 0.000\;130(2) \nonumber
  \;,
\end{eqnarray}
where the uncertainty comes from inaccuracy of the numerical calculations of $F_{WK}(30)=0.0020$ \cite{john}, which is
estimated here as a value of a unit in the last digital place. Comparing the results above one can
find a conservative estimate: $F_{WK}(7)=0.000\;134 (6)$. The value $F_{WK}(30)=0.0020$ is valid for both the $1s$ and $2s$
states and we use this value for the $3s$ state as well. On this level of accuracy  
($\delta F_{WK}(30)\simeq 0.0001$)
there is no shift of the $2p$ levels \cite{john} and we use a zero value for them.

\subsection{Fitting of one-loop self energy contributions}

We separate from the expression for the self-energy part of the one-loop
correction \eq{exp} the function
\begin{equation} \label{fitlow}
  G(Z) = A_{60} + \langle \mbox{\rm higher-order terms} \rangle\;.
\end{equation}
Using numerical values of $A_{60}$ from Refs. \cite{pach93,jent96}
and ones of $G(Z)$ from Refs. \cite{mohr922,jent99}, we
performed several types of fitting for these functions.

We started with fitting {\it (I)} with function
\begin{equation}
\label{fitIa}
 A_{60} +  (Z\alpha) \left(A_{71} \ln\frac{1}{(Z\alpha)^2} +
 A_{70}\right)
 \,,
\end{equation}
minimizing the sum
\begin{equation}
\label{fitIb}
 \sum_{Z}{
 \left\{
 \frac{
  \widetilde G(Z)
  - \left[
  A_{60}
  + (Z\alpha) \left(A_{71}\ln\bigl(1/(Z\alpha)^2\bigr) + A_{70}\right)
  \right]}
 { \delta G(Z) }
 \right\}^2}
 \,,
\end{equation}
with respect to $A_{70}$ and $A_{71}$ for $1s$ state
(where $Z=1\dots5$) and to $A_{70}$ for $2s$ state
(where $Z=5,10$). In the latter case we used the fact that
$A_{71}(1s)=A_{71}(2s)$.
The statistical error of data
\begin{equation}
 \left[\delta \widetilde G(Z)\right]^2
 = \left[\delta_{\rm num} G(Z)\right]^2
 + \left[\pi^2 (Z\alpha)^3 A^{(0)}_{70}\right]^2
 \,.
\end{equation}
contains uncertainty of numerical integrations in Refs.
\cite{mohr922,jent99} and of
the fit in \eq{fitlow} due to neglecting of
higher-order terms of absolute order $\alpha(Z\alpha)^8m$
(where $A^{(0)}_{70}$ is a result of preliminary fitting
with $\delta \widetilde G = \delta_{\rm num} \widetilde G$ ).

To estimate the additional systematic uncertainty which
originates from the unknown term of order
$\alpha(Z\alpha)^7$ we studied a sensitivity of the fit {\it (I)}
to introduction of some perturbation function $h(z)$
\begin{equation}
  \widetilde G(Z) = G(Z) + h(Z)
  \,,
\end{equation}
The final uncertainty of the fit was calculated as a random sum of
differences of the fits without function $h$ and with
$h(Z)$ from the binomial expansion of the expression
\begin{equation}
\label{uncer1}
  (Z\alpha)^2 \left( \ln \frac{1}{(Z\alpha)^2} + \pi \right)^3
\end{equation}
for the $1s$ and $2s$ states and
\begin{equation}
\label{uncer2}
  (Z\alpha)^2 \left( \ln \frac{1}{(Z\alpha)^2} + \pi \right)^2
\end{equation}
for the $2p$-states and for the difference $G_{2s}-G_{ns}$ (see below).

The logarithm $\ln\bigl(1/(Z\alpha)^2\bigr)$ in the expansion \eq{exp}
is a large value\footnote{Note that a natural value for the constant term
is about $ \ln\big(k_0(ns)\big)\sim 3$).} at very low $Z$ but it is
quite a smooth function of $Z$ around $Z=7$
(see Table~\ref{TabLog}).  Due to that, we can also use
a non-logarithmic fitting function
\begin{equation}
\label{II_III}
 G(Z_0) + A (Z-Z_0) + \frac{B}{2}(Z-Z_0)^2
\end{equation}
with smooth behaviour at $Z\sim 7$. In particular, we applied the Eq. (\ref{II_III})
for numerical data from Ref. \cite{mohr922,jent99} at $Z=3,4,5$ with central value $Z_0=4$
(fit {\it II}\/) and $Z=5,10,15$
with $Z_0=10$ (fit {\it III}\/).

\begin{table}
\caption{Function $\ln\bigl(1/(Z\alpha)^2\bigr)$ for small $Z$}
\label{TabLog}
\begin{center}
\def\arraystretch{1.4}
\setlength\tabcolsep{5pt}
\begin{tabular}{cc@{\qquad\qquad}cc}
\hline
$Z$ & $\ln\bigl(1/(Z\alpha)^2\bigr)$ & $Z$ & $\ln\bigl(1/(Z\alpha)^2\bigr)$ %\vphantom{$\Biggm|$} 
\\
\hline
 1 & 9.840 &  7 & 5.949 \\
 2 & 8.454 &  8 & 5.682 \\
 3 & 7.643 &  9 & 5.446 \\
 4 & 7.068 & 10 & 5.235 \\
 5 & 6.622 & 15 & 4.424 \\
 6 & 6.257 & 20 & 3.849 \\
\hline
\end{tabular}
\end{center}
\end{table}

For the $2s$ state we also performed independent fits for $G_{1s}$
and the difference $G_{1s}-G_{2s}$, finding $G_{2s}$ as their
combination. Values of corresponding fits are labeled {\it (IV)}
(fits for low $Z$), {\it (V)} ($Z=3,4,5$ for $1s$ and $Z=5,10,15$
for the difference $G_{1s}-G_{2s}$) and {\it (VI)} (both fits for $Z=5,10,15$).
That can be useful because data on the $1s$ state is more accurate \cite{mohr922,jent99},
and in case of difference the uncertainty is smaller (cf. Eqs. (\ref{uncer1})
and (\ref{uncer2})) and one
of higher-order parameters is known ($A_{71}=0$) \cite{jetp94,zp97}.

Only few data are available for $3s$ \cite{mohr921} at $Z=10,20,30\dots$ and we perform two
fitting: fit {\it I} with $Z=10,20$ and fit {\it III} with $Z=10,20,30$ at $Z_0=20$.

\begin{figure}[th]
\begin{center}
\begin{minipage}{0.45\textwidth}
{\includegraphics[width=\textwidth,bb=68 200 527 630]{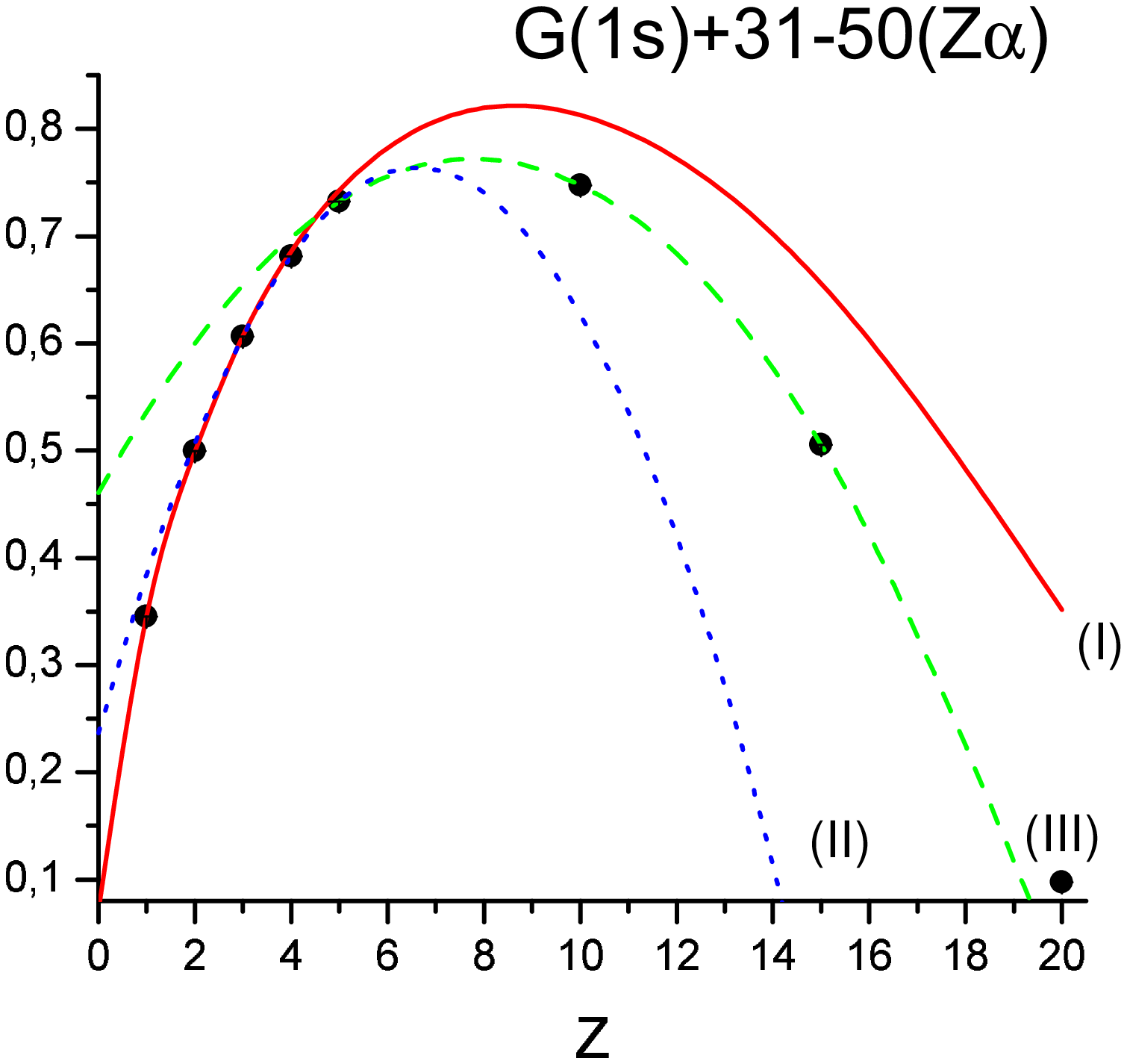}}
\end{minipage}%
\hskip 0.1\textwidth
\begin{minipage}{0.45\textwidth}
{\includegraphics[width=\textwidth,bb=68 200 527 630]{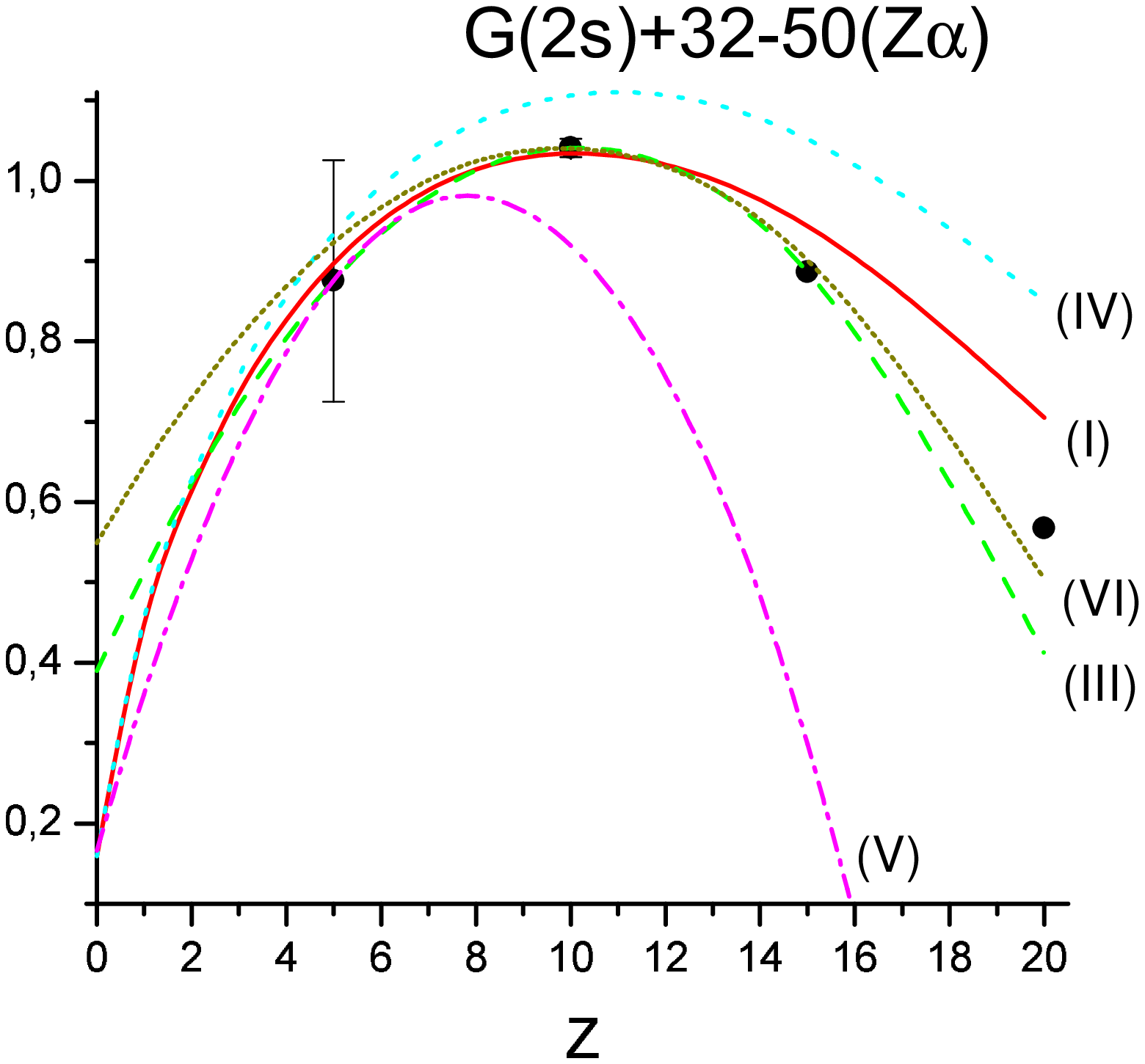}}
\end{minipage}
\end{center}
\caption{Fitting of $G(Z)$ for 1s and 2s states.}
\label{Fig1}
\end{figure}

\begin{figure}[th]
\begin{center}
\begin{minipage}{0.45\textwidth}
{\includegraphics[width=\textwidth,bb=68 200 527 630]{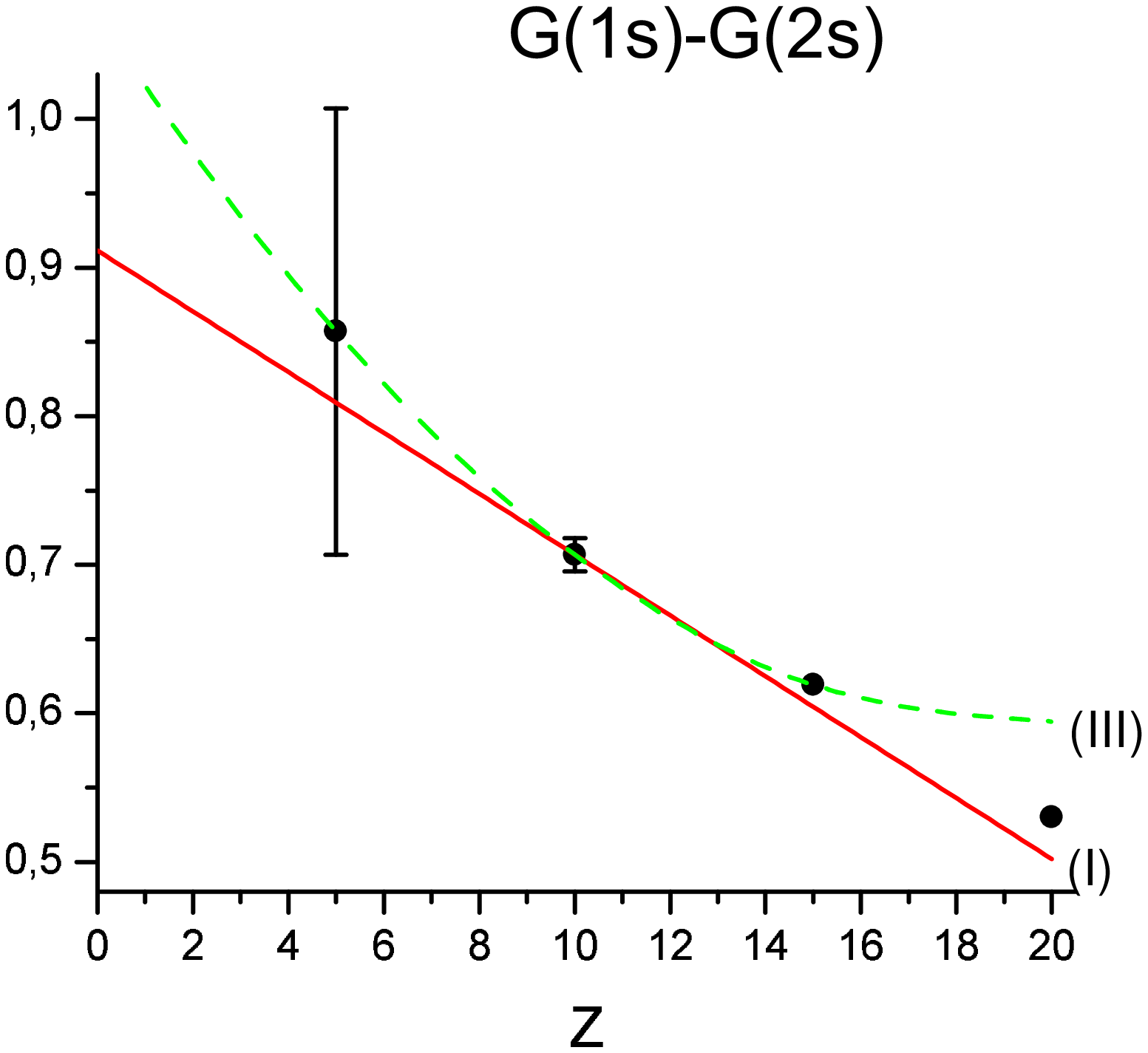}}
\end{minipage}%
\hskip 0.1\textwidth
\begin{minipage}{0.45\textwidth}
{\includegraphics[width=\textwidth,bb=68 200 527 630]{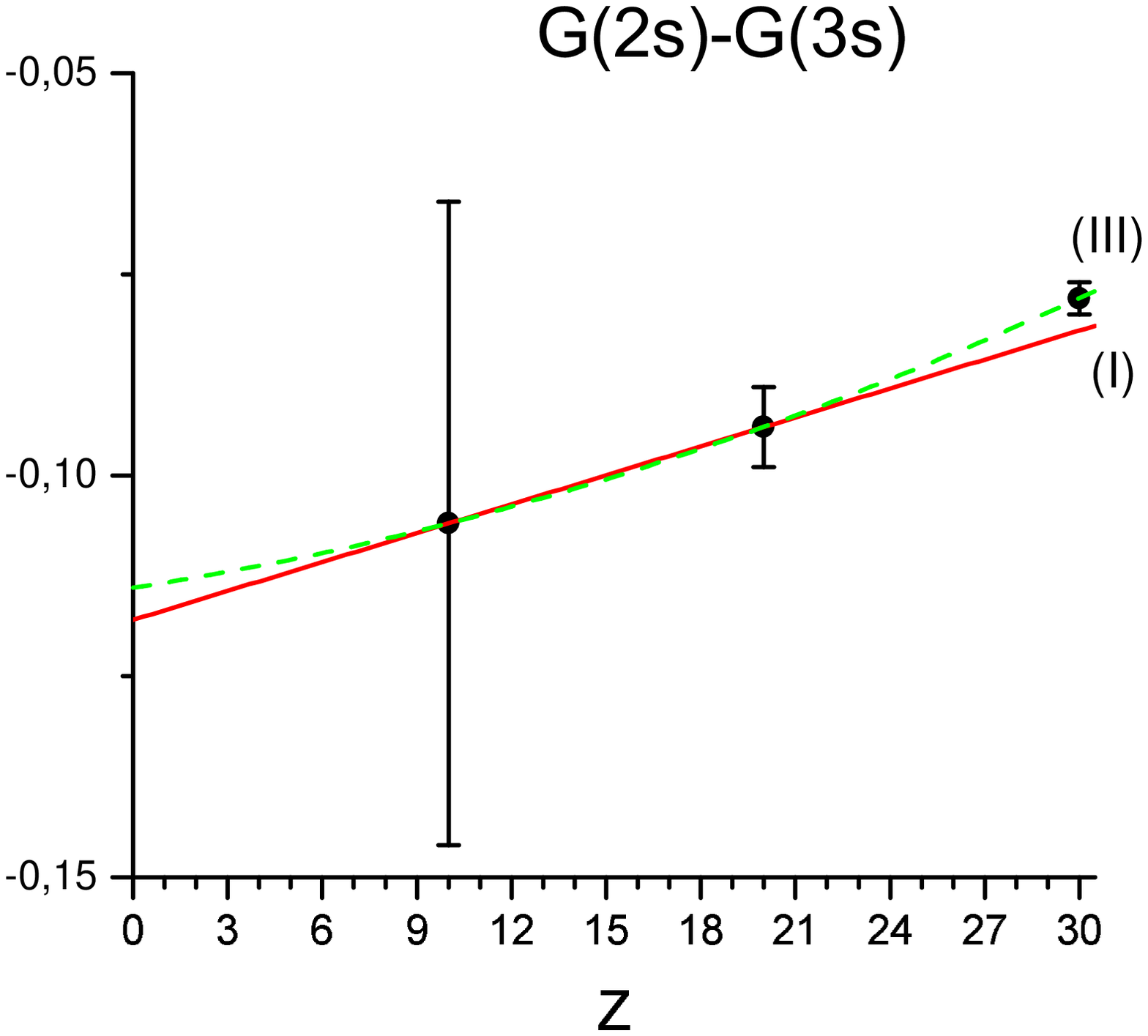}}
\end{minipage}
\end{center}
\caption{Fitting of $G_{1s}(Z)-G_{2s}(Z)$ and $G_{2s}(Z)-G_{3s}(Z)$.}
\label{Fig2}
\end{figure}

\begin{figure}[th]
\begin{center}
\begin{minipage}{0.45\textwidth}
{\includegraphics[width=\textwidth,bb=68 200 527 630]{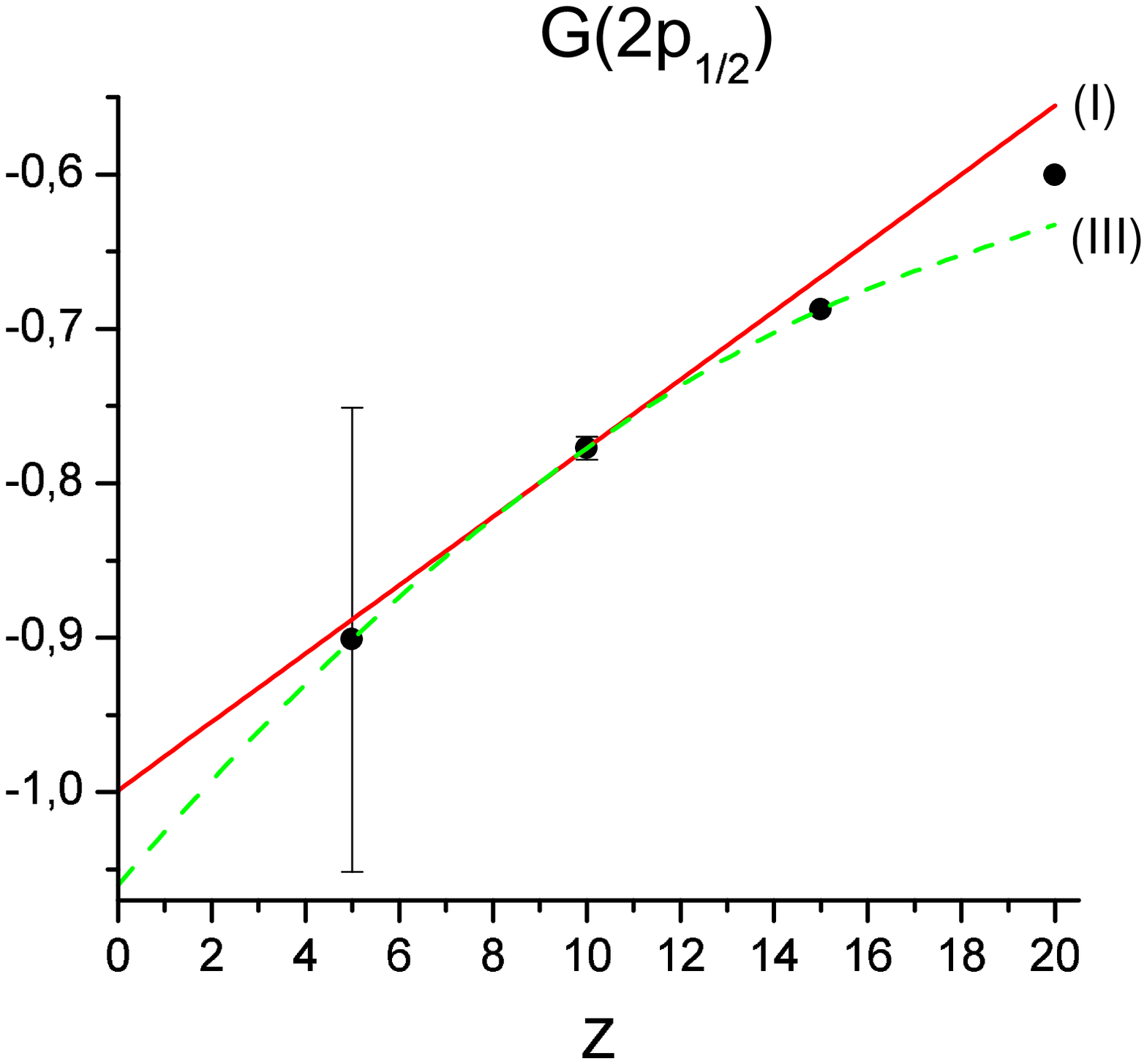}}
\end{minipage}%
\hskip 0.1\textwidth
\begin{minipage}{0.45\textwidth}
{\includegraphics[width=\textwidth,bb=68 200 527 630]{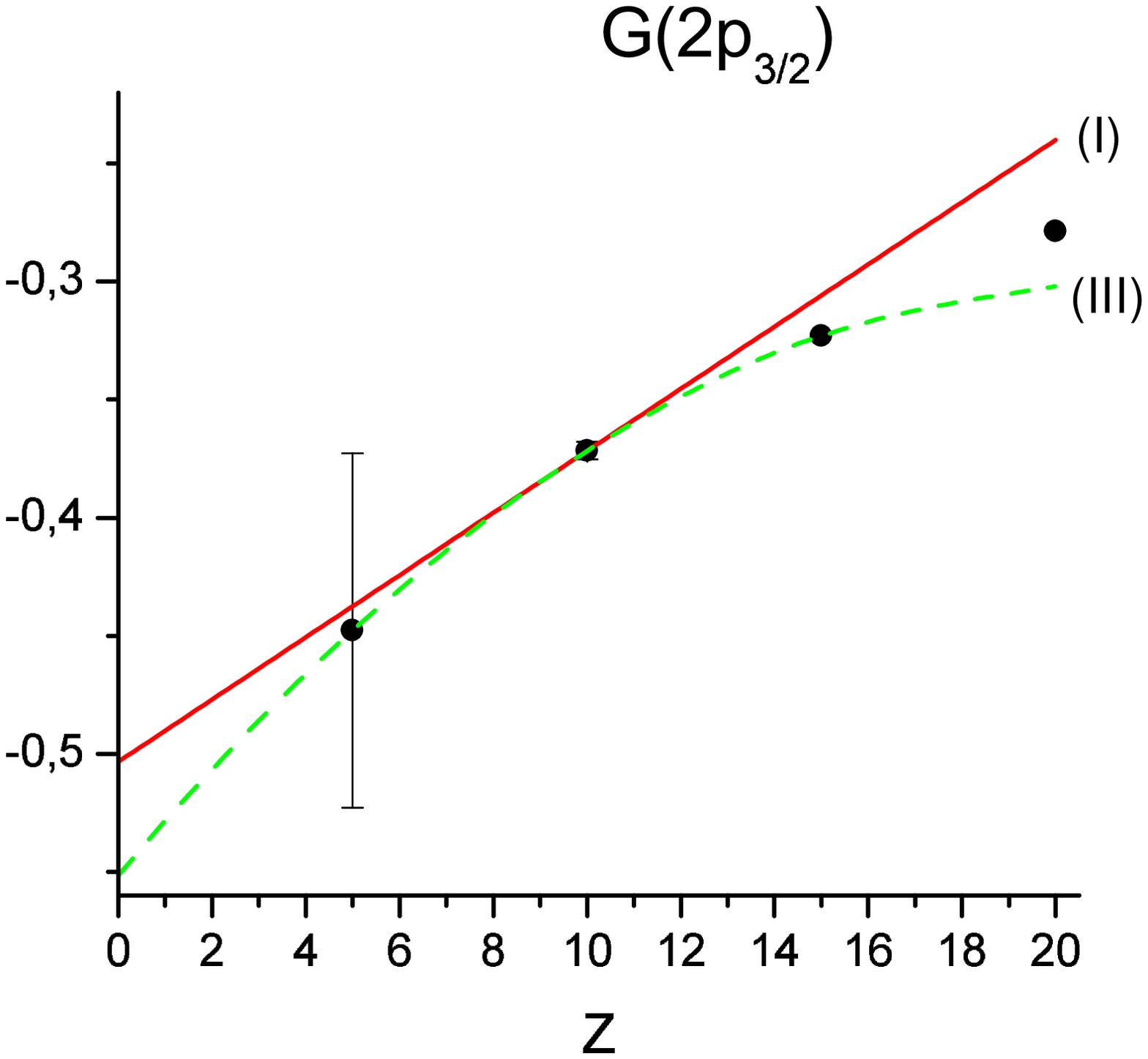}}
\end{minipage}
\end{center}
\caption{Fitting of $G(Z)$ for the 2p$_{1/2}$ and 2p$_{3/2}$ states.}
\label{Fig3}
\end{figure}

Different fitting functions are plotted in Figs. \ref{Fig1}--\ref{Fig3} for
$n=1$, $n=2$, $G_{1s}-G_{2s}$ and $G_{2s}-G_{3s}$. The points with error bars are for
numerical values obtained
in Refs. \cite{mohr921,mohr922,jent99}.

The results of all fits for $Z=2$ are summarized in Table~\ref{TabG2} and
$Z=7$  in Table~\ref{TabG7}.
The value for the $1s$ state at $Z=7$ is in agreement with
the less accurate one obtained in Ref. \cite{yer}.

\begin{table}
\caption{Values $G(2)$ for different types of fitting}
\label{TabG2}
\begin{center}
\def\arraystretch{1.4}
\setlength\tabcolsep{5pt}
\begin{tabular}{lcc}
\hline
Value
 & {\it (I)}
 & {\it (IV)} \\
\hline
$G_{1s}$        & -29.772(7) &  \\
$G_{2s}$        & -30.7(2)   & -30.64(3)  \\
$G_{1s}-G_{2s}$ & 0.87(3)    &   \\
$G_{2s}-G_{3s}$ & -0.12(9)    &  \\
$G_{2p_{1/2}}$     & -0.95(3)    & \\
$G_{2p_{3/2}}$     & -0.48(3)    & \\
\hline
\end{tabular}
\end{center}
\end{table}

\begin{table}
\caption{Values $G(7)$ for different types of fitting}
\label{TabG7}
\begin{center}
\def\arraystretch{1.4}
\setlength\tabcolsep{5pt}
\begin{tabular}{lcccccc}
\hline
Value
 & {\it (I)}
 & {\it (II)}
 & {\it (III)}
 & {\it (IV)}
 & {\it (V)}
 & {\it (VI)} \\
\hline
$G_{1s}$       & -27.6(1) & -27.68(1) & -27.677(9) &&& \\
$G_{2s}$       & -28.5(1) &           & -28.47(5)  & -28.4(1)  & -28.47(5) & -28.45(3) \\
$G_{1s}-G_{2s}$& 0.77(3)  &           & 0.79(5) &   &  &  \\
$G_{2p1/2}$ & -0.84(3) &           & -0.85(5)   &&& \\
$G_{2p3/2}$ & -0.41(3) &           & -0.41(3)   &&& \\
\hline
\end{tabular}
\end{center}
\end{table}

\subsection{Two-loop contributions}

Two-loop corrections have not yet been calculated exactly and only a few of the terms
have been known up-to-date \cite{two5,jetp93}:
\begin{eqnarray}
 H_{ns}(Z) =&~&
\left( - \frac{2179}{648}-\frac{10}{27}\pi^2
 + \frac{3}{2}\pi^2\ln(2)
 - \frac{9}{4}\zeta(3)\right)  -   21.556(3) \, (Z\alpha)
\nonumber \\
&+& (Z\alpha)^2 \left[ - \frac{8}{27} \, \ln^3 \frac{1}{(Z\alpha)^2 } + G^{II}_{ns}(Z)\right]
 \,, \\
 H_{2p_{1/2}}(Z) =& - &\frac{1}{3}\, a_4
 + (Z\alpha)^2 \left[ \frac{1}{9}  \,\ln^2 \frac{1}{(Z\alpha)^2 } + G^{II}_{2p_{1/2}}(Z)\right]
 \,, \\
 H_{2p_{3/2}}(Z) =& ~& \frac{1}{6}\, a_4
 + (Z\alpha)^2 \left[ \frac{1}{9} \, \ln^2 \frac{1}{(Z\alpha)^2 }+ G^{II}_{2p_{1/2}}(Z)\right]
 \,,
\end{eqnarray}
where for $l\neq 0$ the leading contribution is due to the anomalous magnetic moment of
the electron\footnote{We follow a notation
$(g_e-2)_{\rm QED} = 2\cdot \sum{a_{2n}\,\big(\alpha/\pi\big)^n}$.
See Ref. \cite{kinoshita} for detail.}
\begin{equation}
 a_4 =
 \frac{197}{144}+\frac{\pi^2}{12}-\frac{\pi^2}{2}\ln(2)+\frac{3}{4}\zeta(3)
 \simeq -0.328\,478\,966\dots
 \,.
\end{equation}
The higher-order terms denoted as $G^{II}(Z)$ are not known. We expect \cite{karshenboim99} that 
the magnitude of $G^{II}$ does not exceeded by half the value of the leading logarithmic term ($\ln^3(Z\alpha)$ for the $ns$ states and 
$\ln^2(Z\alpha)$ for the $2p$ states) at $Z=1$ and use the estimation for $G^{II}(1)$ for any $Z$. 
The purpose of this calculation is to eliminate any other sources of theoretical uncertainty and 
to study a value of $G^{II}_{2s}$ by a comparison with experiment.

The leading state-dependent term for the $s$-state is also known \cite{jpb,zp97}, and
in particular one can find
\begin{eqnarray}
 H_{1s}(Z) - H_{2s}(Z) &=& (Z\alpha)^2 \left[\ln^2 \frac{1}{(Z\alpha)^2}
 \left( \frac{16}{9} \ln(2) - \frac{7}{3} \right)+ G^{II}_{12}(Z)\right]\,,
 \\
 H_{2s}(Z) - H_{3s}(Z) &=& (Z\alpha)^2 \left[\ln^2 \frac{1}{(Z\alpha)^2}
 \left( \frac{16}{9} \ln \frac{3}{2} - \frac{91}{81} \right)+ G^{II}_{23}(Z)\right]\,,
\end{eqnarray}
and unknown higher-order contributions $G^{II}$ are estimated in the same way as for the $2p$ states.

\subsection{Three-loop contributions}

The three-loop contributions for $ns$ state at $Z=0$ was eventually obtained
in Ref. \cite{meln}
\begin{equation}
 K_{ns}(0) = 0.4174\dots
 \,,
\end{equation}
while for the $p$-states the corrections
\begin{eqnarray}
 K_{2p_{1/2}}(0) &=& -\frac{1}{3} \, a_6 \,, \\
 K_{2p_{3/2}}(0) &=&  \frac{1}{6} \, a_6 \,,
\end{eqnarray}
comes from the $g\!-\!2$ of electron:
\begin{equation}
 a_6 = 1.181\,241\dots
\end{equation}

\subsection{Pure recoil corrections}

The pure recoil correction
\begin{equation}
 \Delta E_{\rm Rec}
 =
 \frac{1}{\pi}\,\frac{1}{n^3}\,\frac{m^2}{M} (Z\alpha)^4 R(Z)\; ,
\end{equation}
where is known analytically with sufficient accuracy \cite{recoil}
\begin{eqnarray}
 R_{ns}(Z) =&~&
 (Z\alpha) \left[
  \frac{2}{3} \ln \frac{1}{(Z\alpha)}
  - \frac{8}{3} \ln\big(k_0(ns)\big) + \frac{187}{18}
 \right] \nonumber\\
 &+&
 (Z\alpha)^2 \, \pi \left( 4 \ln(2)- \frac{7}{2} \right)
 \,,\\
 R_{2p}(Z) =&~&
 (Z\alpha) \, \left[ - \frac{8}{3} \ln\big(k_0(2p)\big)  - \frac{7}{18} \right]
 + (Z\alpha)^2 \, \frac{\pi}{3}\;.
\end{eqnarray}
Some numerical results for the states with $n=1,2$ are also available \cite{numrec}.
Second-order recoil corrections ($(Z\alpha)^4(m^3/M^2)$) are known \cite{recoil2}
but their contribution is negligible.

\subsection{Radiative-recoil corrections}

The radiative-recoil corrections are known only in the leading order
\cite{pach952}
\begin{equation}
  \Delta E_{\rm RRC}(nl) =
  \frac{\alpha(Z\alpha)^5}{\pi}\,\frac{m}{n^3}\,\frac{m}{M}\,
  \left(-1.36449\right)\,\delta_{l0}
  \,.
\end{equation}

\subsection{Finite-nuclear-size correction}

The correction for $s$-states due to the finite size of the nucleus is of the form
\begin{equation}
\label{NSlead}
 \Delta E_{\rm Nucl} (nl) =
 \frac{2}{3}\frac{(Z\alpha)^4\,m}{n^3} \, \Big(m\, R_{\rm N}\Big)^2\,\delta_{l0}\,
 \left( 1 +
 (Z \alpha)^2 \ln \frac{1}{Z\alpha m \, R_{\rm N}}
 \right)
 \,,
\end{equation}
where $R_N$ is the rms nuclear charge distribution radius. The leading term in Eq.~(\ref{NSlead}) vanishes
for the $p$ states because their non-relativistic wave function is vanishes at the origin itself.
However, the small component of the Dirac wave function contains a factor of ${\bf \sigma}{\bf p}/2m$
and the Dirac wave function is not equal to zero at the origin. In particular, one can easily find
\begin{eqnarray}
\Bigl(\Psi_{2p_{1/2}}(0)\Bigr)^2 &=& {3\over 32}  \frac{(Z\alpha)^5 m^3}{\pi}\;,
\nonumber\\
\Bigl(\Psi_{2p_{3/2}}(0)\Bigr)^2 &=& 0\;,
\end{eqnarray}
and
\begin{eqnarray}
\Delta E_{\rm Nucl} (2p_{1/2}) &=&
{1\over 16}\,
(Z\alpha)^6 \, m \, \Big(m\, R_{\rm N}\Big)^2\,\;,
\nonumber\\
\Delta E_{\rm Nucl} (2p_{3/2}) &=&0\;.
\end{eqnarray}
The error of the logarithmic term can be estimated as
$ 1/2 \, \Delta E_{\rm Nucl} \, (Z \alpha)^2 \ln(1/\alpha)$.
The nuclear radii and corrections are presented in Table~\ref{TabRN}.

\begin{table}
\caption{Nuclear square charge radius}
\label{TabRN}
\begin{center}
\def\arraystretch{1.4}
\setlength\tabcolsep{5pt}
\begin{tabular}{cccc}
\hline
Atom & $R_{\rm N}$ & $\Delta E_{\rm Nucl}(2s)$ & Ref. \\
& [fm] & [MHz] &  \\
\hline
${}^{4}{\rm He}$ & 1.674(12) & 8.80(12) & \protect{\cite{RHe4}}\\
${}^{14}{\rm N}$ & 2.560(11) & 3145(33) & \protect{\cite{RN14}}\\
${}^{15}{\rm N}$ & 2.612(9)  & 3274(30) & \protect{\cite{RN15}}\\
\hline
\end{tabular}
\end{center}
\end{table}

\section{Summary}

All contributions and final values of the Lamb shifts in the lowest states
of hydrog-like ions $^4$He$^{+}$, ${}^{14}$N$^{6+}$ and ${}^{15}$N$^{6+}$ are presented in
Tables~\ref{TabHe}, \ref{TabN14} and~\ref{TabN15}, respectively.
The final results for intervals which can be measured
are listed in Table~\ref{TabSplit}.
The results in Table~\ref{TabSplit} involve three main sources of uncertainty
and we split the uncertainty there and in auxiliary
Tables~\ref{TabHe}--\ref{TabN15}, respectively:
\begin{itemize}
\item The higher-order two-loop corrections;
\item Nuclear structure corrections (They require further study);
\item Other theoretical QED uncertainties beyond the higher-oder two-loop effects.
\end{itemize}
The theory (with unknown higher-order two-loop effects {\em excluded\/})
is found to be accurate enough and we hope the study of helium and nitrogen
hydrogen-like ions is a promising way to study in detail
the two-loop contributions experimentally.

\begin{table}
\caption{Different contributions to the Lamb shift in hydrogen-like helium ${}^{4}{\rm He}$
(in MHz)}
\label{TabHe}
\begin{center}
\def\arraystretch{1.4}
\setlength\tabcolsep{5pt}
\begin{tabular}{lr@{}lr@{}lr@{}l}
\hline
Contribution & \multicolumn{2}{c}{$2s$} & \multicolumn{2}{c}{$2p_{1/2}$} &
\multicolumn{2}{c}{$2p_{3/2}$} \\
\hline
SE               & 14\,251&.51(2) & -204&.747(10)  & 201&.503(10)   \\
VP               & -426&.78       & -0&.022        & -0&.005      \\
WK               & 0&.02          & 0&             & 0&           \\
\hline
One-loop         & 13\,824&.75(2) & -204&.769(10) & 201&.498(10) \\
Two-loop         & 0&.70(11)      & 0&.420(2)     & -0&.201(2) \\
Three-loop       & 0&.004         & 0&.003        & 0&.002     \\
Pure recoil      & 2&.53          & -0&.131       & -0&.131     \\
Radiative recoil & -0&.01         & 0&            &  0&        \\
\hline
QED              & 13\,827&.98(11)(2)     & -204&.484(2)(10) & 201&.168(2)(10) \\
Nuclear size     & 8&.80(13)              & 0&               & 0&         \\
\hline
Total            & 13\,836&.77(11)(13)(2) & -204&.484(2)(10) & 201&.168(2)(10) \\
\hline
\end{tabular}
\end{center}
\end{table}

\begin{table}
\caption{Different contributions to the Lamb shift in ${}^{14}$N (in MHz)}
\label{TabN14}
\begin{center}
\def\arraystretch{1.4}
\setlength\tabcolsep{5pt}
\begin{tabular}{lr@{}lr@{}lr@{}l}
\hline
Contribution & \multicolumn{2}{c}{$2s$} & \multicolumn{2}{c}{$2p_{1/2}$} &
\multicolumn{2}{c}{$2p_{3/2}$} \\
\hline
SE              & 1\,390\,601&(19) &  -29\,298&(19) & 31\,088&(19) \\
VP              & -62\,021&        & -40&           & -8&        \\
WK              & 32&              & 0&             & 0&           \\
\hline
One-loop        & 1\,328\,611&(19) & -29\,338& & 31\,080&  \\
Two-loop        & -411&(209)       & 68&(2)    & -25&(2) \\
Three-loop      & 0&.5             & -0&.5     & 0&.3    \\
Pure recoil      & 319&            & -17&      & -17&    \\
Radiative recoil & -2&             & 0&        &  0&       \\
\hline
QED              & 1\,328\,517&(209)(19) & -29\,287&(2)(19) & 31\,038&(2)(19) \\
Nuclear size     & 3\,145&(33)           & 6&               & 0& \\
\hline
Total            & 1\,331\,662&(209)(33)(19) & -29\,282&(2)(19) & 31\,038&(2)(19) \\
\hline
\end{tabular}
\end{center}
\end{table}

\begin{table}
\caption{Different contributions to the Lamb shift in ${}^{15}$N (in MHz)}
\label{TabN15}
\begin{center}
\def\arraystretch{1.4}
\setlength\tabcolsep{5pt}
\begin{tabular}{lr@{}lr@{}lr@{}l}
\hline
Contribution & \multicolumn{2}{c}{$2s$} & \multicolumn{2}{c}{$2p_{1/2}$} &
\multicolumn{2}{c}{$2p_{3/2}$} \\
\hline
SE                 & 1\,390\,611&(19) &  -29\,298&(19) & 31\,088&(19) \\
VP                 & -62\,022&        & -40&           & -8&        \\
WK                 & 32&              & 0&             & 0&           \\
\hline
One-loop          & 1\,328\,620&(19) & -29\,338& & 31\,080&  \\
Two-loop          & -411&(209)       & 68&(2)    & -25&(2) \\
Three-loop        & 0&.5             & -0&.5     & 0&.3    \\
Pure recoil       & 297&             & -16&      & -16&    \\
Radiative recoil  & -2&              & 0&        &  0&       \\
\hline
QED               & 1\,328\,505&(209)(19)  & -29\,286&(2)(19)  & 31\,039&(2)(19) \\
Nuclear size      & 3\,274&(30)            & 6&                & 0& \\
\hline
Total             &1\,331\,779&(209)(30)(19) & -29\,280&(2)(19) & 31\,039&(2)(19) \\
\hline
\end{tabular}
\end{center}
\end{table}

\begin{table}
\caption{Differences of energies (in MHz)}
\label{TabSplit}
\begin{center}
\def\arraystretch{1.4}
\setlength\tabcolsep{5pt}
\begin{tabular}{ccc}
\hline
Atom & Value & Result \\
\hline
${}^{4}{\rm He}$ & $\Delta E(2s-2p_{1/2})$ & 14\,041.25(11)(13)(2)       \\
${}^{4}{\rm He}$ & $8E_L(2s)-27E_L(3s)$ & -768.35(29)        \\
${}^{14}{\rm N}$ & $\Delta E(2p_{3/2}-2s)$ & 25\,030\,522(209)(33)(27) \\
${}^{15}{\rm N}$ & $\Delta E(2p_{3/2}-2s)$ & 25\,030\,475(209)(30)(27) \\
\hline
\end{tabular}
\end{center}
\end{table}

\section*{Acknowledgments}

We are grateful to Ed Myers for stimulating discussions. The work was supported in part
by RFBR (grant 00-02-16718) and
Russian State Program ``Fundamental Metrology''. A support (SK) by a NATO grant CRG 960003 is
also acknowledged.

\end{document}